\documentclass[11pt]{article}
\usepackage{moriond,epsfig}

\def\be{\begin{equation}}
\def\ee{\end{equation}}
\def\bea{\begin{eqnarray}}
\def\eea{\end{eqnarray}}

\def\vet{\vec E_T}
\def\met{\not\!\!\vet}
\def\gev{~\mathrm{GeV}}

\begin{document}
\vspace*{4cm}
\title{SEARCHES FOR A HIGH MASS STANDARD MODEL HIGGS BOSON AT THE TEVATRON}

\author{ S. PAGAN GRISO\\(for CDF and D\O\ Collaborations) }

\address{INFN and University of Padova,\\Department of Physics G. Galilei, Via Marzolo 8,35131 Padova, Italy}

\maketitle

\abstracts{Higgs boson searches are commonly considered one of the main objectives of particle physics nowadays. The latest results obtained by the CDF and D\O\ collaborations are presented here when seatching for Higgs boson decaying into a W-boson pair, currently the most sensitive channel for masses greater than $130~\mathrm{GeV}$. The presented results are based on an integrated luminosity that ranges from $3.0$ to $4.2~\mathrm{fb}^{-1}$. No significant excess over expected background is observed and the $95\%$ CL limits are set for a Standard Model (SM) Higgs boson for different mass hypotheses ranging from $100~\mathrm{GeV}$ to $200~\mathrm{GeV}$. The combination of CDF and D\O\ results is also presented, which exclude for the first time a SM Higgs boson in the $160 < m_{H} < 170~\mathrm{GeV}$ mass range. 
}

\section{Higgs searches at the Tevatron}

 Hadron colliders collision data can be used to probe Standard Model (SM) theory in several sectors. Data collected at the $p\bar p$ Tevatron collider operating at $\sqrt{s}=1.96~\mathrm{TeV}$ has been useful for both re-enstablish or improve previous important measurements and to make new important discoveries, looking for processes whose cross section range by several orders of magnitude. CDF and D\O\ experiments\cite{CDFDetector}~\cite{DZeroDetector} are collecting data since 2002 and nowadays they have around $5.4~\mathrm{fb}^{-1}$ of good data available for analysis. The Higgs searches presented here are based on an integrated luminosity that goes from $3.0$ to $4.2~\mathrm{fb}^{-1}$ depending on the specific analysis.
 
 According to SM, Higgs boson is produced at Tevatron energies by four main production mechanisms, for a total cross-section of about $0.6~\mathrm{pb}$\footnote{Higgs boson properties depend on its mass, which is not predicted inside the SM framework; here and whenever not explicitly mentioned in the following, numbers refer as an example to an Higgs boson of mass $m_{H}=160~\mathrm{GeV}$.}. The dominant mechanism is via gluon fusion and fermionic loop ($\sim 86\%$), then there is the production with a $W$ or $Z$ boson ($\sim 15\%$) and finally by vector-boson fusion (VBF).

 SM Higgs decay depends on its mass $m_H$. For $m_H > 135\gev$ the main decay channel is a W-boson pair, and searches of this final state at Tevatron are commonly defined as {\it High Mass Higgs Searches}; if $m_H<135\gev$ $H\rightarrow b\bar b$ predominantly and these searches are described in another proceeding to this conference.

 CDF and D\O\ look for both Ws decaying leptonically, selecting events with two electrons or two muons or one electron and one muon for a total branching ratio of the $WW$ pair of $\sim 6\%$, including the leptonic decay of the $\tau$. The larger branching ratio dacay in hadrons is not used due to high level of QCD background. The di-lepton final state offer a clean signature at hadron colliders and the trigger cross section is under control also at high instantaneous luminosity.

 The main background consists of Drell-Yan (DY) events, which have the same final visible state of the signal (two leptons) but no real missing energy and can be reduced requiring it. Heavy di-boson production ($WW$, $WZ$, $ZZ$) is the most important not reducible background for these searches. $t\bar t$ is also an important background, in particular for events containing jets. Finally instrumental backgrounds arise from $W/Z+\gamma$ or $jets$ events where the photon or the jet is misidentified as a lepton. 
 Most of these processes are modelled using PYTHIA Monte Carlo and a GEANT-based simulation of the detectors. An important exception is the $WW$ background: CDF models NLO effects using a pure NLO simulation, namely MC@NLO
, while D\O\ uses Sherpa
to model the $p_T$ of the $WW$ system\cite{sherpastudy}. These predictions are then normalized to NNLO cross section calculations for $WH, ZH, t\bar t$ processes, NLO for $VBF$, $WW$, $WZ$, $ZZ$, $W\gamma$. 
 The gluon fusion signal process has been simulated using the most recent calculation available\cite{FlorianGrazziniggH}
which uses the recent MSTW2008 parton density functions (pdf) set\cite{MSTW2008}.
 Data-driven methods are used in order to model the constribution of the instrumental background.

\section{Analysis description}

 To select signal events both collaborations reuire two high-$p_T$ opposite sign isolated leptons. 
In order to increase acceptance to Higgs events, dedicated analyses also look for final state containing two leptons with the same charge; they will be briefly discussed in section~\ref{SSana}. 
CDF requires the first (second) lepton to have $p_T$ greater than $20$ ($10$)$\gev$, while D\O\ asks both leptons to have $p_T$ ($E_T$) greater than $10$ ($15$)$\gev$ for muons (electrons). A significant transverse missing energy $\met$ is then required to reduce DY. The invariant mass of the lepton pair must be greater than $16(15)\gev$ at CDF(D\O) in order to suppress heavy flavor decays and DY events.

 Table \ref{tab:yieldspresel} shows the numbers of events expected and observed after the pre-selection cuts. The signal/background ratio is too low for a counting experiment to discriminate background-only from signal plus background hypotheses. Other kinematical properties are used to enhance the separation.
The strongest discriminant is the opening angle between final state leptons; in signal events leptons come from the decay of spin-1 particles which are from spin-0 Higgs boson, which implies, given the fixed helicity for neutrinos, that leptons will tend to go in the same direction.
For the background this spin correlation does not exist and final state leptons tend to be back-to-back. To exploit this and other kinematical signal properties both collaborations use multivariate techniques. An Artificial Neural Network (NN) is trained to separate signal from background for each different Higgs mass hypothesis. Moreover, in order to exploit different signal and background composition, the selected sample is divided in several analysis channels.

\begin{table}[th]
\caption{Number of expected and observed events after pre-selections for CDF and D\O\ $H\rightarrow WW$ opposite-sign analyses. D\O\ Numbers include statistical uncertainties only.\label{tab:yieldspresel}}
\vspace{0.4cm}
\begin{center}
\begin{tabular}{|l|c|cccc|}
\multicolumn{5}{c}{\tiny Tevatron Preliminary $\int \mathcal{L} = 3.6-4.2 \; \rm{fb}^{-1}$, $M_H=160$ GeV} \\
\hline
 & $\mathcal{L}$ (fb$^{-1}$) & Signal & Background & $S/\sqrt{B}$ & Data \\
\hline
CDF & 3.6 & $20.0\pm 2.5$ & $1088\pm105$ & $0.61$ & $1085$\\
D\O\ (stat. only)  & 3.0-4.2 &$23.2\pm 0.1$ & $4994\pm 30$ & $0.33$ & $4749$ \\
\hline
\end{tabular}
\end{center}
\end{table}


\subsection{D\O\ analysis}

 D\O\ collaboration separate the analysis
depending on the flavor of the final state leptons: $ee$, $e\mu$, $\mu\mu$, with a collected integrated luminosity of, respectively,  $4.2$, $4.2$ and $3.0~\mathrm{fb}^{-1}$.
 Input to the NNs can be classified in three different types: lepton specific variables (e.g. $p_T$ of the leptons), kinematic properties of the whole event (e.g. $\met$) or angular variables (e.g. $\Delta\phi(leptons)$).

 No significan excess over predicted background is observed in D\O\ data and 95\% CL upper limits on the production cross section of a SM Higgs boson are set using a modified frequentist approach ($CL_s$)\cite{CDFDZeroCombinationPaper}, which are $70\%$ higher than the predicted SM cross section for $m_H=160\gev$\cite{D0HWWPubNote}.


\subsection{CDF analysis}

 CDF colaboration separate the $3.6~\mathrm{fb}^{-1}$ selected sample depending on jet multiplicity, optimizing different NNs for each sample. Jets are requested to have $E_T>15\gev, \vert \eta\vert < 2.5$. Input variables to the NNs are analogous to D\O.
 Events with no jets have signal contribution only from gluon fusion and the dominant background is $WW$ production. Events with one jet have additional signal contribution from associate Higgs production and VBF and $WW$ still remain the main source of background. Finally, signal events with two or more jets are dominated by $WH$,$ZH$ and $VBF$ production mechanisms and the dominant background comes from $t\bar t$.

 No significant excess over predicted background is observed in CDF data and $95\%$ CL upper limit on the production cross section of a SM Higgs boson are set using a Bayesian technique\cite{CDFDZeroCombinationPaper}, which for $m_H=160~GeV$ are $50\%$ higher than the predicted SM cross section\cite{CDFHWWPubNote}.


\subsection{Same sign analyses}
\label{SSana}

 In order to increase the acceptance to Higgs events, both collaborations also perform searches with two same charge final state leptons. The main signal contribution comes from $WH\rightarrow WWW\rightarrow l^{\pm}l^{\pm}+X$, where one of the leptons comes from the W boson produced in association with the Higgs. 
 For these searches the main backgrounds are instrumental backgorunds coming from charge mis-identification or jets faking a lepton signature. 
 CDF uses $3.6~\mathrm{fb}^{-1}$ and an analysis technnique similar to the opposite sign analysis to set $95\%$ CL upper limits on the production of a SM Higgs boson that are $6.6$ times the expected SM cross section\cite{CDFHWWPubNote} ($\sigma^H_{SM}$).
 D\O\ recently\footnote{This result is not currently included in the Tevatron combination discussed in section \ref{sec:tevcomb}} made public a new result which use $3.6~\mathrm{fb}^{-1}$ of data and set limits up to $18.4\cdot\sigma^H_{SM}$\cite{D0SSPubNote}.

\section{Tevatron combination and results}
\label{sec:tevcomb}

 Results of both collaborations, in the High and Low mass region, are combined using two different methods: a Bayesian and a modified frequentist technique\cite{CDFDZeroCombinationPaper}. Both perform a counting experiment for each bin of the final discriminant, including effects from systematics uncertainties. 

 This combination procedure is able to correlate systematics uncertainties among different analysis and experiments. Systematics uncertainties are divided in two main categories: rate and shape systematics. The first ones affect the normalization of the different signal and background contributions; these are the most important and are dominated by theoretical uncertainties on signal and background cross sections used to normalize our simulations. Shape systematics affect the shape of the output discriminant; an important example is the jet energy scale calibration.

\begin{figure}
\begin{center}
\psfig{figure=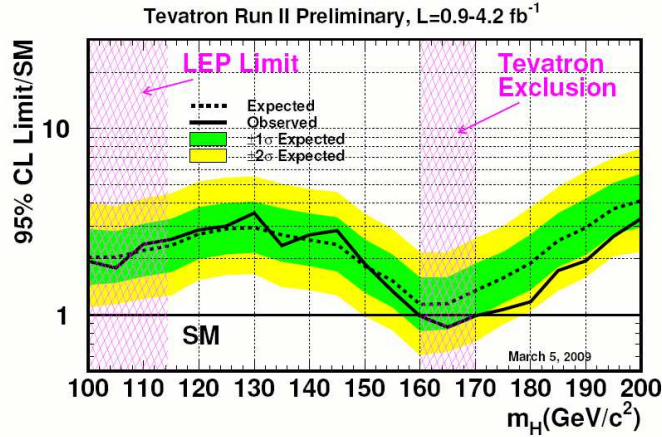,height=2.3in}
\end{center}
\caption{Observed and expected (median, for the background-only hypothesis) $95\%$ CL limits on the ratios to the SM cross section, as functions of the Higgs boson mass for the combined CDF and D\O\ analyses.
\label{fig:TevatronLimits}}
\end{figure}

 Figure \ref{fig:TevatronLimits} summarize the combination for Higgs masses between $100$ and $200\gev/c^{2}$.
 The dotted line represent the median, the green and yellow bands one and two sigma spread of the distribution of the expected limits from a background-only hypothesis; the solid line is the limit that is set looking at data. Each analysis is performed for different Higgs mass hypotheses in $5\gev$ steps, then results are connected with a straight line for better readability. 
The combination {\bf excludes at 95\% CL a Standrd Model Higgs boson in the \mbox{\boldmath$160 < m_H < 170\gev/\mathrm{c}^2$}} mass range; the expected limits are $1.1$ and $1.4$ times the expected SM cross section respectively for an Higgs mass of $160$ and $170\gev$.

\section{Conclusions}
 During latest years CDF and D\O\ have improved their sensitivity to low cross section processes, been now in the reaches of the Higgs production. The combination of the analysis carried out by the two experiments has led for the first time to the exclusion of a SM Higgs in the $160-170\gev$ mass range. More data is available to be analyzed and more will be collected in the next years allowing an exclusion by each experiment and widening the combined exclusion region.

%


\section*{Acknowledgments}
I thank organizers and who joined this conference for the stimulating discussions that I'll bear in mind. A special thank to D. Lucchesi for her illuminating guidance in preparing my contribution.

\end{document}